\documentstyle[12pt,a4wide]{article}

\begin{document}

\author{B. Boisseau\thanks{E-mail: boisseau@celfi.phys.univ-tours.fr}
 and B. Linet\thanks{E-mail: linet@celfi.phys.univ-tours.fr} \\
\small Laboratoire de Math\'ematiques et Physique Th\'eorique \\
\small CNRS/UPRES-A 6083, Universit\'e Fran\c{c}ois Rabelais \\ 
\small Facult\'e des Sciences et Techniques, Parc de Grandmont 
37200 TOURS, France }

\title{ \bf Dynamics of a self-gravitating thin string \\
in scalar-tensor theories of gravitation}

\date{}

\maketitle

\begin{abstract}
We examine the dynamics of a self-gravitating string in the scalar-tensor 
theories of gravitation by considering a thin tube of matter to describe it.
For a class of solutions, we obtain in the generic case that the extrinsic 
curvature of the world sheet of the central line is null in the limit where
the radius of the string tends to zero. However, if we impose a specific 
constraint on the behaviour
of the solution then we find that only the mean curvature of the world
sheet of the central line vanishes which is just the Nambu-Goto
dynamics. This analysis can include the massless dilatonic theories of gravity.

{\em PACS numbers: 11.27, 04.25}
\end{abstract}

\thispagestyle{empty}

\section{Introduction}

The knowledge of the equations of motion of the cosmic strings is essential
to predict the evolution and the 
physical consequences of an eventual string network in the
early Universe \cite{vil,bra,hin}. In general, the Nambu-Goto dynamics of
the cosmic strings is postulated in a background curved spacetime. 
Yet, there is no proof that a self-gravitating 
cosmic string obeys
the Nambu-Goto dynamics as the strings in field theory moving
in the Minkowski spacetime \cite{nie,for}. On the contrary, several 
authors \cite{vic,fro,unr,cla} state
in general relativity that a world sheet whose the points are conical
singularities is totally geodesic which is a very particular case of a
minimal surface characteristic of the history of a Nambu-Goto string.

However, a recent paper \cite{boi} showed in general relativity the
existence of self-gravitating thin strings of matter having their central
line governed by the Nambu-Goto dynamics in the limit where their radius 
goes to zero. 

The purpose of this work is to examine the 
dynamics of a self-gravitating 
cosmic string in the scalar-tensor theories of gravitation 
\cite{ber,wag,nor} which
generalize the Brans-Dicke theory \cite{dic}. There is a renewed
interest of these theories of gravitation due to the importance of the 
scalar field or dilaton in the low-energy effective string theory 
presently used to study the early Universe. In general
the dilaton is massive and consequently the theory is practically
the general relativity for distances much larger than the range of the
dilaton. However, a massless dilaton would be possible \cite{dam2} and in this
case our method can be immediately transposed.

The straight U(1) gauge cosmic string has been studied in the Brans-Dicke
theory \cite{gun}, in the scalar-tensor theories \cite{gui} 
and in the dilatonic theories of gravity \cite{gre}. 
Now the distribution of matter vanishes practically beyond the range $l_{0}$ 
of the massive
gauge and Higgs fields, therefore we can substitute to the U(1) gauge
cosmic string a tube of matter describing a cylinder of radius $l_{0}$
having as exterior solution
the vacuum solution which coincides with the asymptotic
solution of the U(1) gauge cosmic string.

Then, for a thin string describing a tube of matter of arbitrary shape, we 
adopt the assumption that the exterior metric is locally the one describing 
a straight string. The interior solution is constrained from this. The aim
of the work is to find the equations of motion
of the central line of the thin tube of matter in the limit where its
radius goes to zero.

The plan of the work is as follows. In section 2, we recall the basic
equations of the scalar-tensor theories and the dilatonic theory. We give
in section 3 the form of the interior and exterior metric representing a
static cylinder of matter. In section 4, we introduce a self-gravitating
thin string of arbitrary shape. We derive the equations of motion in
section 5 by taking the limit where its radius tends to zero. We add
in section 6 some concluding remarks.

\section{Gravitational field equations}   

The gravitational field variables of the scalar-tensor theories are the metric
$\hat{g}_{\mu \nu}$ of the spacetime and the scalar field $\phi$. 
Nevertheless, it is more convenient to use a non-physical
metric $g_{\mu \nu}$ defined by $g_{\mu \nu}=A^{-2}(\phi )\hat{g}_{\mu \nu}$
where the function $A(\phi )$ characterizes each scalar-tensor theory. 
The matter fields $\psi_{M}$ are coupled as usual to the metric 
$\hat{g}_{\mu \nu}$ within the Lagrangian density
${\cal L}_{M}[\psi_{M},\hat{g}_{\mu \nu}]$. The field equations can be 
then derived from the action
\begin{equation}
\label{s1}
S=\frac{1}{16\pi {\cal G}}\int d^{4}x\sqrt{-g}\left( R-2g^{\mu \nu}
\partial_{\mu}\phi \partial_{\nu}\phi \right) +\int d^{4}x
{\cal L}_{M}[\psi_{M},A^{2}(\phi )g_{\mu \nu}]
\end{equation}
where ${\cal G}$ is the coupling gravitational constant which can be related
to the Newtonian constant. By varying action (\ref{s1}) with respect to 
$g_{\mu \nu}$ and $\phi$, we get the gravitational field equations 
\begin{equation}
\label{1}
R_{\mu \nu}-2\partial_{\mu}\phi \partial_{\nu}\phi =8\pi {\cal G}
(T_{\mu \nu}-\frac{1}{2}Tg_{\mu \nu}) \, ,
\end{equation}
\begin{equation}
\label{2}
\Box \phi =-4\pi {\cal G}\alpha (\phi )T  \quad
{\rm with} \quad \alpha (\phi )=\frac{1}{A}\frac{dA}{d\phi}
\end{equation}
and by varying with respect to $\psi_{M}$ the matter field equations. 
The source $T_{\mu \nu}$ is related to the conserved energy-momentum tensor
$\hat{T}_{\mu \nu}$ by $T_{\mu \nu}=A^{2}(\phi )\hat{T}_{\mu \nu}$; it 
satisfies the equation
\begin{equation}
\label{s2}
\nabla_{\mu}T_{\nu}^{\mu}=\alpha (\phi )T\partial_{\nu}\phi \, .
\end{equation}

We emphasize that we cannot give independently $g_{\mu \nu}$ and $\phi$ to
deduce from them the source $T_{\mu \nu}$. We have necessarly the condition
\begin{equation}
\label{s3}
R-2g^{\mu \nu}\partial_{\mu}\phi \partial_{\nu}\phi =
\frac{2}{\alpha (\phi )}\Box \phi
\end{equation}
resulting from the trace of field equations (\ref{1}). We now turn to the 
inverse.
We consider that $g_{\mu \nu}$ and $\phi$ verify (\ref{s3}). We can calculate
$T_{\mu \nu}$ by (\ref{1}) and then (\ref{2}) is automatically verified. 
Moreover it is easy to see that the source $T_{\mu \nu}$ satisfies (\ref{s2}).

We now consider the massless dilatonic theory of gravity. The field 
variables are the string metric $\hat{g}_{\mu \nu}$, the dilaton $\phi$ and
the matter field $\psi_{M}$. The metric $g_{\mu \nu}$ in the Einstein 
frame is defined by $g_{\mu \nu}=\exp (-2\phi )\hat{g}_{\mu \nu}$. 
Following Damour and Polyakov \cite{dam2}, the field equations can be derived
from the following action
\begin{equation}
\label{s4}
S=\frac{1}{16\pi \overline{G}}\int d^{4}x \sqrt{-g} \left( R-
2g^{\mu \nu}\partial_{\mu}\phi \partial_{\nu}\phi \right)+\int d^{4}x
{\rm e}^{-2\phi}{\cal L}_{M}[\psi_{M},{\rm e}^{2\phi}g_{\mu \nu}] \, .
\end{equation}
in the simplest case. We obtain the gravitational field equations
\begin{equation}
\label{s5}
R_{\mu \nu}-2\partial_{\mu}\phi \partial_{\nu}\phi =
8\pi \overline{G}(T_{\mu \nu}-\frac{1}{2}Tg_{\mu \nu}) \, ,
\end{equation}
\begin{equation}
\label{s6}
\Box \phi =-4\pi \overline{G}\sigma \quad {\rm with} \quad
\sigma =\frac{1}{\sqrt{-g}}\frac{\delta ({\rm e}^{-2\phi}{\cal L}_{M})}
{\delta \phi}
\end{equation}
where $\overline{G}$ is a bare constant coupling   
and also the field matter equations. The quantity
$\sigma$ is not directly related to the $T_{\mu \nu}$; it depends on
the matter field $\psi_{M}$ as $T_{\mu \nu}$. We point out that the
physical energy-momentum tensor $T_{\mu \nu}$ is not conserved since we have
\begin{equation}
\label{s7}
\nabla_{\mu}T_{\nu}^{\mu}=\sigma \partial_{\nu}\phi \, .
\end{equation}
As in the previous case, we cannot give independently $g_{\mu \nu}$ and
$\phi$ but the condition takes into account the matter field.

\section{Static, cylindrically symmetric solutions}

The vacuum solutions to field equations (\ref{1}) and (\ref{2}) with
$T_{\mu \nu}=0$ which are static and cylindrically symmetric are known as
solutions dependent only on one coordinate \cite{rub}.
By adding the boost symmetry, they have finally the general form
\begin{equation}
\label{3}
ds^{2}=\left( \frac{\rho -\overline{l}_{0}}{\rho_{*}}\right)^{k}
(-dt^{2}+dz^{2})+d\rho^{2}+B^{2}(\rho -\overline{l}_{0})^{2}
\left( \frac{\rho -\overline{l}_{0}}{\rho_{*}}\right)^{-2k}d\varphi^{2} \, ,
\end{equation}
\begin{equation}
\label{4}
\phi =\phi_{0}+\kappa \ln 
\left(\frac{\rho -\overline{l}_{0}}{\rho_{*}}\right) \quad
{\rm with} \quad \kappa =\sqrt{k\left( 1-\frac{3}{4}k\right)}
\end{equation}
in the coordinate system $(t,z,\rho ,\varphi )$ with $0\leq \varphi <2\pi$.
For $k=0$, we recover the metric with conical singularity of the general
relativity. For $k\neq 0$, the curvature tensor of 
metric (\ref{3}) blow up at $\rho =\overline{l}_{0}$.  
We can perform a redefinition of $\rho_{*}$ and 
a rescaling of $t$ and $z$ to set $B=1$. Metric (\ref{3}) depends only
on two arbitrary constants.  

Metric (\ref{3}) and scalar field (\ref{4}) are
respectively the asymptotic metric and the asymptotic scalar field of the
solution for a straight U(1) gauge cosmic string. 
Then, the constants appearing in expression 
(\ref{3}) can be approximatively determined in this case \cite{gun,gui,gre}. 
The lengths $\rho_{*}$ and $l_{0}-\overline{l}_{0}$ have the
same order of magnitude as the range  $l_{0}$ of the gauge and Higgs fields..  
We emphasize that $k$ is quadratic in ${\cal G}$ and consequently
very small but strictly positive.

We substitute a tube of matter representing a cylinder in place of this 
U(1) gauge cosmic string. Its energy-momentum tensor
will be determined by the data 
of the interior metric via the gravitational field equations.
The interior metric is regular at the
central line $\rho =0$ and it matches to the exterior metric (\ref{3})
at $\rho =l_{0}$, likewise the scalar field. In the next sections, we will
perform a limit process in which $l_{0}$ tends to zero but the
constants $B$ and $k$ keep the same values and in which $\rho_{*}$ 
and $l_{0}-\overline{l}_{0}$ are dependent on 
$l_{0}$. To this end, we put the following family
of regular metrics parametrized by $\epsilon$
\begin{equation}
\label{5}
ds^{2}=\left\{  \begin{array}{ll}
X\left( \displaystyle{\rho \strut \over \epsilon} \right) 
(-dt^{2}+dz^{2})+d\rho^{2}+
\epsilon^{2}h^{2}\left( \displaystyle{\rho \strut \over \epsilon}\right) 
\varphi^{2} & 0\leq \rho <l_{0} \\
\left( \displaystyle{ \rho -\overline{l}_{0}\strut \over \rho_{*}}\right)^{k}
(-dt^{2}+dz^{2})+d\rho^{2}+B^{2}(\rho -\overline{l}_{0})^{2}
\left( \displaystyle{\rho -\overline{l}_{0}\strut \over \rho_{*}}\right)^{-2k}
d\varphi^{2} & \rho >l_{0} \end{array} \right. 
\end{equation}
where the assumption of regularity at $\rho =0$ is ensured by the behaviours
\[
X(x)=1+X_{2}x^{2}+\cdots \quad {\and} \quad {\rm and} \quad 
h(x)=x+h_{3}x^{3}+\cdots \, .
\]
The requirement $X(0)=1$  in (\ref{5}) 
fixes the rescaling of the coordinates $t$ and $z$
and consequently it justifies to keep the constant $B$. The matching 
conditions at $\rho =l_{0}$
require the continuity of the components of the metric and their first
derivative with respect to $\rho$. After some manipulation, we get
\begin{eqnarray}
\label{7}
\nonumber & & l_{0}-\overline{l}_{0}=(1-k)\epsilon 
\frac{h(l_{0}/\epsilon )}{h'(l_{0}/\epsilon )} \, , \\
& & B=\frac{1}{1-k}h'\left( \frac{l_{0}}{\epsilon} \right) \left[ (1-k)
\frac{\epsilon}{\rho_{*}}
\frac{h(l_{0}/\epsilon )}{h'(l_{0}/\epsilon )}\right]^{k} \, , \\
\nonumber & & X\left(\frac{l_{0}}{\epsilon}\right) =
\left( \frac{l_{0}-\overline{l}_{0}}
{\rho_{*}}\right)^{k} \, , \quad 
X'\left(\frac{l_{0}}{\epsilon}\right) =k\frac{\epsilon}{\rho_{*}}
\left(\frac{l_{0}-\overline{l}_{0}}{\rho_{*}}\right)^{k-1} \, .
\end{eqnarray}

It is easy to show that a change of 
parameter $\epsilon$ in form (\ref{5}) corresponds to a 
redefinition of the functions $h$ and $X$ which is compatible with (\ref{7}).  
If $\epsilon_{1}$ and $\epsilon_{2}$
denote two parameters, we have 
$\epsilon_{1}h_{1}(\rho /\epsilon_{1})=\epsilon_{2}h_{2}(\rho /\epsilon_{2})$
and $X_{1}(\rho /\epsilon_{1})=X_{2}(\rho /\epsilon_{2})$. 
Thus, we can choose the parameter $\epsilon$ of the family of metrics (\ref{5})
to be
\begin{equation}
\label{8} 
\epsilon =\rho_{*} \, .
\end{equation}
In general relativity, another geometrical choice had been made \cite{boi}.

For the scalar field, we choose the following form
\begin{equation}
\label{9}
\phi =\left\{ \begin{array}{ll}
\phi_{0}+Z\left(\displaystyle{\rho \strut \over  \epsilon} \right) 
& 0\leq \rho <l_{0} \\
 \phi_{0}+\kappa \ln \left( \displaystyle{\rho -\overline{l}_{0} \strut
\over \epsilon}\right)  
& \rho >l_{0} \end{array} 
\right.
\end{equation}
taking account (\ref{8}) and with the behaviour
\[
Z(x)=Z_{2}x^{2}+Z_{4}x^{4}+\cdots
\]
to ensure the regularity at $\rho =0$.
Scalar field (\ref{9}) and its first derivative with respect to $\rho$ 
are continuous at $\rho =l_{0}$; we get
\begin{equation}
\label{10}
Z\left(\frac{l_{0}}{\epsilon}\right) 
=\kappa \ln \left( \frac{l_{0}-\overline{l}_{0}}{\epsilon}\right)  \quad
{\rm and} \quad Z'\left(\frac{l_{0}}{\epsilon}\right) 
=\kappa \frac{\epsilon}{l_{0}-\overline{l}_{0}} \, .
\end{equation}

In summarizing, it is possible to find a family of metrics (\ref{5}) and
scalar fields (\ref{9}) representing a cylinder of matter of radius $l_{0}$.
The exterior solution is characterized by fixed constants $B$ and $k$. Now we
can take $l_{0}\rightarrow 0$ but without considering a line source
which is a difficult concept in non-linear field equations \cite{ger}. 
For given functions $h$ and $X$, we see from (\ref{7}) that the ratio
$l_{0}/\epsilon$ remains constant. It is equivalent to take
$\epsilon \rightarrow 0$ in the family of solutions (\ref{5}) and (\ref{9}).

When we will use these solutions we will furthermore need to the continuity of
the second derivative with respect to $\rho$ of the 
components of metric (\ref{5}) and scalar field (\ref{9}). It is not
necessary to write down them. We confine ourselves to these solutions. 
In consequence, $T_{\mu \nu}$ vanishes at $\rho =l_{0}$. 

\section{A model of self-gravitating thin string}

We now consider a general tube of matter with a circular section of
finite radius $l_{0}$ whose the central line has an arbitrary shape. 
The central line sweeps a timelike world sheet. A
local coordinate system $(\tau^{A},\rho^{a})$ can be then  attached to the
spacetime by taking the two coordinates $\tau^{A}$ of the world sheet and
the two geodesic coordinates $\rho^{a}$ pointing in a direction
orthogonal to the world sheet. The polar coordinates $\rho$ and 
$\varphi$ are related to $\rho^{a}$ by
$\rho^{1}=\rho \cos \varphi$ and $\rho^{2}=\rho \sin \varphi$. The world
sheet is represented by the equation $\rho =0$.

We imagine that this cylinder of matter describes a cosmic string. By 
assuming that the typical variation on which we have
a change in the direction of the central line is much larger than the 
radius $l_{0}$, we can suppose
that its metric can  approximatively be identified with metric (\ref{5}) 
in a small neighbourhood.
We insist that the constants $B$ and $k$ have the same
values during all the history of the tube of matter. We restrict ourselves to
metric having no cross terms and so we put in the coordinates 
$(\tau^{A},\rho ,\varphi )$
\begin{equation}   
\label{11}
ds^{2}=\left\{ \begin{array}{ll}
X\left( \displaystyle{\rho \strut \over \epsilon} \right)
g_{AB}(\tau^{C},\rho ,\varphi )d\tau^{A}d\tau^{B}+
d\rho^{2}+\epsilon^{2}h^{2}
\left( \displaystyle{\rho \strut \over \epsilon} \right) d\varphi^{2}
& 0\leq \rho <l_{0} \\
\left(\displaystyle{\rho -\overline{l}_{0}\strut \over \epsilon} 
\right)^{k} g_{AB}(\tau^{C},\rho , \varphi )
d\tau^{A}d\tau^{B}+ & \\
\quad \quad d\rho^{2}+B^{2}(\rho -\overline{l}_{0})^{2}
\left(\displaystyle{\rho -\overline{l}_{0}\strut \over \epsilon} \right)^{-2k}
d\varphi^{2} & \rho >l_{0} \end{array} 
\right.
\end{equation}
where the typical variation of components $g_{AB}$ is much larger than
$\epsilon$. The matching conditions at $\rho =l_{0}$ are given by 
(\ref{7}) as in the previous section. Instead of polar coordinates,
we prefer express metric (\ref{11}) in coordinates $\rho^{a}$ and we
denote $g_{ab}$ the metric of the surface $\tau^{A} =$ constant. These 
components are functions of $\rho^{a}/\epsilon$. We remark that 
$g_{ab}=\delta_{ab}+O(\rho^{2})$.

According to metric (\ref{11}), the metric of the world sheet is
\begin{equation}
\label{11a}
\gamma_{AB}=g_{AB}\mid_{\rho =0}
\end{equation}
and the extrinsic and mean curvature of the world sheet are
\begin{equation}
\label{11b}
K_{aAB}=\frac{1}{2}\partial_{a}g_{AB}\mid_{\rho =0} \quad {\rm and} \quad
K_{a}=K_{aAB}\gamma^{AB}
\end{equation}
since $X(0)=1$ and $X'(0)=0$.

Further calculations, that we do not reproduce here, show that the scalar 
field $\phi$ of such a solution is very particular. We anticipate and we take
directly
\begin{equation}
\label{12} 
\phi =\left\{ \begin{array}{ll}
\phi_{0}+Z\left(\displaystyle{\rho \strut \over \epsilon} \right) 
& 0\leq \rho <l_{0} \\
\phi_{0}+\kappa \ln \left( \displaystyle{\rho -\overline{l}_{0} \strut \over
\epsilon}\right)
& \rho > l_{0} \end{array} \, .
\right.
\end{equation}
The matching conditions at $\rho =l_{0}$ are given by (\ref{10}).
Since $h$, $X$ and $Z$ are supposed known from
the straight string, we obtain from (\ref{s3}) that the components $g_{AB}$
giving the evolution of the string of arbitrary shape must verify
\begin{equation}
\label{s8}
R=2g^{ab}\partial_{a}Z\partial_{b}Z+\frac{2}{\alpha (\phi_{0}+Z)}
\frac{1}{\sqrt{{\rm det}g_{ab}{\rm det}g_{AB}}X}
\partial_{a}\left( \sqrt{{\rm det}g_{ab}{\rm det}g_{AB}}X
g^{ab}\partial_{b} Z \right) \, .
\end{equation}

We add moreover that the second derivative with respect to $\rho$ of
the metric and the scalar field are continuous. We think that the cylinder
of matter described by metric (\ref{11}) and scalar field (\ref{12}) is
reasonably general in despite of (\ref{s8}) 
and it allows us  to obtain results about the equations of 
motion of the central line having an arbitrary shape.

\section{Equations of motion of the string}

We proceed as in the previous paper \cite{boi}. The method is based on 
an expansion in powers of $1/\epsilon$ of the geometrical quantities  and of 
the scalar field associated to the family of solutions (\ref{11}) and
(\ref{12}). We use the fact that
\[
g_{ab}\mid_{\rho =l_{0}}=O(\frac{1}{\epsilon^{0}}) \, , \quad
\partial_{c}g_{ab}\mid_{\rho =l_{0}}=O(\frac{1}{\epsilon}) \, , \quad
X\mid_{\rho =l_{0}}=O(\frac{1}{\epsilon^{0}}) \, , \quad
\partial_{c}X\mid_{\rho =l_{0}}=O(\frac{1}{\epsilon}) \, , \quad \dots
\]
On the contrary, we use
\[
g_{AB}\mid_{\rho =l_{0}}=O(\frac{1}{\epsilon^{0}}) \, , \quad
\partial_{c}g_{AB}\mid_{\rho =l_{0}}=O(\frac{1}{\epsilon^{0}}) \, , \quad \dots
\]
since the typical variation of $g_{AB}$ is much larger than
$\epsilon$. The variations of the different quantities are thus 
characterized by $\epsilon$. 
We now expand in powers of $1/\epsilon$ the gravitational 
field equations on the circle $\rho =l_{0}$. By our choice of matching 
conditions, we have $T_{\mu \nu}\mid_{\rho =l_{0}}=0$ and therefore we
obtain the following form of the equations
\begin{equation}
\label{13}
R_{\mu \nu}\mid_{\rho =l_{0}}-2\partial_{\mu}\phi 
\partial_{\nu}\phi \mid_{\rho =l_{0}}\equiv L_{\mu \nu}\left(\frac{1}
{\epsilon^{2}}\right) +L_{\mu \nu}\left( \frac{1}{\epsilon}\right) +
L_{\mu \nu}\left( \frac{1}{\epsilon^{0}}\right) =0 \, , 
\end{equation}
\begin{equation}
\label{14}
\frac{1}{\sqrt{-g}}\partial_{\alpha}\left( \sqrt{-g}g^{\alpha \beta}
\partial_{\beta}\phi \right) \mid_{\rho =l_{0}}\equiv L\left( \frac{1}
{\epsilon^{2}}\right) +L\left( \frac{1}{\epsilon}\right) +
L\left( \frac{1}{\epsilon^{0}}\right) =0 \, .
\end{equation}

The terms in $1/\epsilon^{2}$ in (\ref{13}) and (\ref{14}) vanish
identically because they contain only $g_{ab}$, $X$ and $Z$ and
their first and second derivatives at $\rho =l_{0}$ which satisfy the 
vacuum field
equations of the straight string (\ref{5}) and (\ref{9}). For instance, we have
\[
L_{AB}\left( \frac{1}{\epsilon^{2}}\right) =-
\frac{1}{2\sqrt{{\rm det}g_{ab}}}\partial_{a}\left( \sqrt{{\rm det}g_{ab}}
g^{ab}\partial_{b}X\right) g_{AB}\mid_{\rho =l_{0}}=0
\]
as we can easily verified it, likewise for $L_{ab}(1/\epsilon^{2})$, 
$L_{aA}(1/\epsilon^{2})$ and $L(1/\epsilon^{2})$.

We now fulfil a limit process in which $\epsilon \rightarrow 0$, i.e. 
$l_{0}\rightarrow 0$, the ratio $l_{0}/\epsilon$ remaining constant. 
To analyse the now remaining $1/\epsilon$ order equations, we point 
out that we have
\[
\gamma_{AB}=\lim_{l_{0}\rightarrow 0}g_{AB}\mid_{\rho =l_{0}} \quad {\rm and}
\quad K_{aAB}=\frac{1}{2}\lim_{l_{0}\rightarrow 0}\partial_{a}g_{AB}
\mid_{\rho =l_{0}}
\]
from (\ref{11a}) and (\ref{11b}) since the typical variation of 
$g_{AB}$ is much larger than $\epsilon$.
The leading terms  in $1/\epsilon$ in (\ref{13}) and (\ref{14}) must go to zero
as $\epsilon$ tends to zero. It is straightforward to carry out the 
calculation of the $(A,B)$ components
\begin{equation}
\label{15}
\lim_{\epsilon \rightarrow 0}\epsilon L_{AB}\left( \frac{1}{\epsilon}\right) =
F_{1}^{b}\left( \frac{\rho^{a}_{0}}{\epsilon}\right) K_{bAB}-
F_{2}^{b}\left( \frac{\rho^{a}_{0}}{\epsilon}\right) \gamma_{AB}K_{b}=0
\end{equation}
where
\begin{eqnarray}
\label{16}
\nonumber & & F_{1}^{b}\left( \frac{\rho^{a}_{0}}{\epsilon}\right) 
=\lim_{\epsilon \rightarrow 0} \epsilon \frac{1}{2}
\left[ -X\left( \partial_{a}g^{ab}+\Gamma^{a}_{ad}g^{db}\right)
-g^{ab}\partial_{a}X \right] \mid_{\rho =l_{0}} \, , \\
& & F_{2}^{b}\left( \frac{\rho^{a}_{0}}{\epsilon}\right) = 
\lim_{\epsilon \rightarrow 0}\epsilon \frac{1}{2}g^{ab}\partial_{a}X
\mid_{\rho =l_{0}} \, ,
\end{eqnarray}
the $(a,b)$ components
\begin{equation}
\label{17}
\lim_{\epsilon \rightarrow 0}\epsilon L_{ab}\left( \frac{1}{\epsilon}\right) =
F_{ab}^{d}\left( \frac{\rho^{a}_{0}}{\epsilon}\right) K_{d}=0
\end{equation}
where
\begin{equation}
\label{18}
F_{ab}^{d}\left( \frac{\rho^{a}_{0}}{\epsilon}\right) =
\lim_{\epsilon \rightarrow 0}\epsilon \left[ \Gamma_{ab}^{d}-\frac{1}{2}
\frac{\partial_{b}X}{X}\delta_{a}^{d}-\frac{1}{2}\frac{\partial_{a}X}{X}
\delta_{b}^{d}\right] \mid_{\rho =l_{0}} \, ,
\end{equation}
the $(a,A)$ components
\begin{equation}
L_{aA}\left( \frac{1}{\epsilon}\right) =0
\end{equation}
and the scalar field equation
\begin{equation}
\label{19}
\lim_{\epsilon \rightarrow 0}\epsilon L\left( \frac{1}{\epsilon}\right) =
F_{3}^{a}\left( \frac{\rho^{a}_{0}}{\epsilon}\right) K_{a}=0
\end{equation}
where
\begin{equation}
F_{3}^{a}\left( \frac{\rho_{0}^{a}}{\epsilon}\right) =
\lim_{\epsilon \rightarrow 0}\epsilon g^{ab}\partial_{b}\phi 
\mid_{\rho =l_{0}} \, .
\end{equation}

The functions $F_{1}^{b}$, $F_{2}^{b}$, $F_{ab}^{c}$ and $F_{3}^{b}$ are 
functions of $\rho_{0}^{a}/\l_{0}$, i.e. they depend only on the polar
angle $\varphi$. They are proportional to $\rho^{a}/l_{0}$.
We now turn to equation (\ref{15}); since it must be 
verified for all polar angles, we have in the generic case
\begin{equation}
\label{16a}
K_{aAB}=0 \, .
\end{equation}
Then, the other equations (\ref{17}) and (\ref{19}) are obviously satisfied.
So, the world sheet swept by the central line of the string is totally
geodesic.

However, another possibility exists as in general relativity  \cite{boi}. 
We can explicitly calculate $F_{1}^{b}$; we get
\begin{equation}
\label{20}
F_{1}^{b}=-\frac{\rho_{0}^{b}}{2l_{0}}\left[ 
\frac{l_{0}}{\epsilon}\frac{X}{h^{2}}\left( \frac{\epsilon}{l_{0}}
h\, h'-1\right) +X'\right]
\end{equation}
for $h$, $h'$ and $X$, $X'$ at $l_{0}/\epsilon$. By imposing the following 
constraint on the functions $h$ and $X$ at the value $l_{0}/\epsilon$
\begin{equation}
\label{21}
h\left( \frac{l_{0}}{\epsilon}\right)  h'\left( \frac{l_{0}}{\epsilon}\right)
+h^{2}
\left( \frac{l_{0}}{\epsilon}\right) 
\frac{X'(l_{0}/\epsilon )}{X(l_{0}/\epsilon )}=\frac{l_{0}}{\epsilon} \, ,
\end{equation}
we annihilate the coefficient of the extrinsic curvature in (\ref{15}). 
So, equation (\ref{16a}) disappears and the remaining equations give
\begin{equation}
\label{22}
K_{a}=0 \, .
\end{equation}
This expresses that the world sheet of the central line 
is minimal which is the behaviour of the Nambu-Goto string.
We point out that constraint (\ref{21}) is invariant under the choice
of parametrization $\epsilon$ of the family of solutions (\ref{11})
and (\ref{12}). 

In order to obtain the Nambu-Goto dynamics, we must find an interior metric
of the straight string whose the functions $h$ and $X$ satisfy the
matching conditions (\ref{7}), the second order junction conditions
and constraint (\ref{21}) at $\rho =l_{0}$. 
This corresponds to specify the behaviour of the matter.

\section{Conclusion}

In the problem of the equations of motion of a self-gravitating thin cosmic
string in the scalar-tensor theories of gravitation, we have found that 
in generic case the world sheet of the central line is totally geodesic 
i.e. $K_{aAB}=0$ for the metric $g_{\mu \nu}$ and in consequence for
the physical metric $\hat{g}_{\mu \nu}$ since the first derivative of the
scalar field vanishes at $\rho =0$.
However by imposing constraint (\ref{21}) 
on the behavior of metric (\ref{11}), 
the scalar field being given by (\ref{12}),
we have seen that only the mean curvature 
$K_{a}$ vanishes, characteristic of the Nambu-Goto dynamics. 
This result coincides with the one in general relativity \cite{boi}.

However, we have obtained this result only for a  class of solutons 
with a particular scalar field.
Instead of form (\ref{11}) of the metric, we have attempted 
more general forms as
\begin{equation}
\label{s9}
ds^{2}=\left\{  \begin{array}{ll}
X\left(\displaystyle{\rho \strut \over \epsilon} \right)
g_{AB}(\tau^{C},\rho^{c})d\tau^{A}d\tau^{B}
+\theta (\tau^{C},\rho^{c})
g_{ab}d\rho^{a}d\rho^{b} & 0\leq \rho <l_{0} \\
\left(\displaystyle{\rho -\overline{l}_{0}\strut \over \epsilon} \right)^{k}
g_{AB}d\tau^{A}d\tau^{B}+ &  \\
\quad \quad \theta (\tau^{C},\rho^{c})\left(
d\rho^{2}+B^{2}(\rho -\overline{l}_{0})^{2}
\left(\displaystyle{\rho -\overline{l}_{0} \strut \over \epsilon }\right)^{-2k}
d\varphi^{2}\right) & \rho >l_{0}
\end{array} \right.
\end{equation}
but the fact that the field equations yield a scalar field 
independent  on $\tau^{A}$ remains.

To obtain this, we have assumed that the scalar field does not interact
directly with the matter within the framework of
the scalar-tensor theories. Our method can be also applied for the massless
dilatonic theory whose gravitational field equations
(\ref{s5}) and (\ref{s6}) are similar. 
We obtain the same result. In the case of a massive
dilaton, with eventually a specific interaction with the energy per unit
of length for the string \cite{dam}, the problem of the equations of 
motion should be taken up.

In our method for finding the equations of motion, we have replaced the cosmic
string defined in field theory by a tube of matter. An open question
is to determine the equations of motion within the coupled theory of the
field theory and the gravity when the range of the 
massive gauge and Higgs fields tends to zero.

\section*{Acknowledgment}

We would like to thank N. Mohammedi for helpful discussions.

\end{document}